\documentclass[12pt]{amsart}

\newcommand{\CC}{\mathcal C}
\newcommand{\R}{\mathbb R}

\newcommand{\const}{\mathrm{const}}
\newcommand{\PV}{\mathcal{PV}}

\begin{document}
\title[Quantum field theory and quantum mechanics]
{Is quantum field theory a generalization of quantum mechanics?}
\author{A. V. Stoyanovsky}
\address{Russian State University of Humanities}
\email{stoyan@mccme.ru}
\begin{abstract}
We construct a mathematical model analogous to quantum field theory, but without
the notion of vacuum and without measurable physical quantities.
This model is a direct mathematical generalization
of scattering theory in quantum mechanics to path integrals with multidimensional
trajectories (whose mathematical interpretation has been given in a previous paper).
In this model the normal ordering of operators in the Fock space
is replaced by the Weyl--Moyal algebra. This model shows to be useful in proof
of various results in quantum field theory: one first proves these results in the
mathematical model and then ``translates'' them into the usual
language of quantum field theory by more or less ``ugly'' procedures.
\end{abstract}
\maketitle

\section*{Introduction}

The purpose of this paper is to answer the question stated in the title; let us
state it in more detail. Mathematically, it is natural to ask whether one can generalize
the rich and beautiful apparatus of the theory of linear partial differential equations to
multidimensional variational problems, so that the bicharacteristics be replaced by
multidimensional surfaces. This question is closely related with presentation of
solutions of partial differential equations in the form of a path integral. A mathematical
interpretation of the notion of path integral has been given in [1]. One can ask
whether the theory of linear PDE's, in particular the scattering theory in quantum mechanics,
can be generalized to path integrals over multidimensional trajectories.
Since the works of Feynman and others, it is conventional to think that such a generalization
is given by quantum field theory. However, when trying to give a mathematical sense to
this statement, one meets difficulties. The main principal problem is that quantum field
theory has no rigorous mathematical sense and, moreover, is self-contradictory.
The logical contradiction is, according to [2], in that particles are considered separately
from the vacuum in which they move when they are sufficiently far from one another.
However, in fact ``particles continuously interact with vacuum as with the
physical medium in which they move'' ([2], p.~139). The logically correct description of
quantum fields should be a synthesis of the notions of particle and field and, therefore,
should exclude the very notion of vacuum, replacing it with quantum field,
the universal form of matter.

Mathematically, this means that one should not use the Fock space. But then one looses
the opportunity to measure physical quantities such as scattering amplitudes. So we are
left with a non-contradictory but non-physical theory. The description of this theory is
the subject of the present paper. The key role in it is played by the infinite
dimensional Weyl--Moyal algebra $W_0$, see its definition in \S2.
See the book [3] or the paper [4] for the detailed
motivation of introducing this algebra instead of the algebra of operators in the Fock space.
In [3,4] it is said that there is a (not everywhere defined) homomorphism from the
algebra $W_0$ to the algebra of operators in the Fock space. However, when computing
matrix elements in the Fock space of operators like $\varphi(x)^4$ from $W_0$, wee see that
these matrix elements are infinite. To make them finite, one should replace $\varphi(x)^4$ by the normally
ordered operator in the Fock space, a procedure which has no mathematical and logical meaning.

Similarly, when proving many results in quantum field theory, one first states them in our
mathematical model and, second, one translates them into the language of quantum field theory,
introducing the Fock space, normal orderings etc.; this translation is an ``ugly''
procedure, but necessary from the physical point of view. This holds for the construction
of $S$-matrix and Green functions in renormalized perturbation theory (this translation
procedure was erroneously omitted in [3,4]), for the Maslov
quantum field theory complex germ (compare its ``mathematical'' counterpart in [3,4]
with the ``physical'' exposition in [5] or [6]), for two-dimensional conformal
field theory (to appear elsewhere), hopefully for ultrasecondary quantization [7],
for pseudodynamical evolution [8] (whose meaning is to be clarified), etc.
Thus, our mathematical model proves to be useful to physics as well as to mathematics.

Let us describe the contents of the present paper. In \S1 a one-dimensional model of
scattering in quantum mechanics is given in a form suitable for generalization to many
dimensions. In \S2 the mathematical model of renormalized perturbation theory for
the Weyl--Moyal algebra is stated on the example of the $\varphi^4$ model in $\R^{3+1}$.

The author is grateful to V. P. Maslov for helpful discussions.

\section{One-dimensional model: scattering theory}

Consider the Schrodinger equation
\begin{equation}
ih\frac{\partial\psi}{\partial t}=\hat H\psi,
\end{equation}
where $\psi=\psi(t,q)(dq_1\ldots dq_n)^{1/2}$ is a half-form (the wave function)
on configuration space, $q=(q_1,\ldots,q_n)$, $H(t,p,q)=H_0(p,q)+V(t,p,q)$ is the classical
Hamiltonian, where $H_0(p,q)$ is quadratic in $p,q$ and independent of $t$, and $V(t,p,q)$ has
compact support in $t$; $\hat H=\hat H(t)$ is the quantum Hamiltonian operator in the Weyl
calculus [9,3,4] with the Weyl symbol $H(t,p,q)$. Let the interval $[T_1,T_2]$ contain the
support of $V$. Denote by $U$ (resp. $U_0$) the evolution operator of equation (1)
(resp. of equation (1) with $H_0$ instead of $H$) from $T_1$ to $T_2$. Denote by $W$ the
algebra of Weyl quantum observables $\hat\Phi(t)$, $\Phi=\Phi(t,p,q)$,
satisfying the Heisenberg equation
\begin{equation}
\frac{d\hat\Phi}{dt}=\frac1{ih}[\hat H_0,\hat\Phi]=\widehat{\{\Phi,H_0\}},
\end{equation}
where $\{\Phi,H_0\}$ is the Poisson bracket (the latter equality in (2) holds,
because $H_0$ is quadratic). Then the operator $S=S(T_1)=U_0^{-1}U$ naturally
belongs to the algebra $W$. We call $S$ the scattering operator.
If $V(t,p,q)=V_0(t,p,q)+j(t)q$ for a vector function $j(t)=(j_k(t))$ with compact support, then
the operator $S=S(j)$ becomes dependent on $j$. We call
\begin{equation}
Tq_{k_1}(t_1)*_{V_0}\ldots*_{V_0}q_{k_N}(t_N)\stackrel{\text{def}}{=}
\left.\frac{\delta^NS(j)}{\delta j_{k_1}(t_1)\ldots\delta j_{k_N}(t_N)}\right|_{j\equiv 0}
\end{equation}
the {\it operator Green function} of equation (1); it is a $W$-valued distribution of
$t_1,\ldots,t_N$. According to [1], there exists a $W$-valued distribution $\mu$
on the space $R$ of trajectories $q(t)$, $-\infty<t<\infty$, such that
\begin{equation}
Tq_{k_1}(t_1)*_{V_0}\ldots*_{V_0}q_{k_N}(t_N)
=\int_R q_{k_1}(t_1)\ldots q_{k_N}(t_N)D\mu(q(\cdot)).
\end{equation}
We call $\mu$ the {\it operator Feynman measure} on the space of trajectories.

The operator $S=S(V)$ satisfies the following conditions:

1) unitarity: $S*\bar S=1$, where $*$ is the (Moyal) product in the Weyl algebra, $\bar S$
has the Weyl symbol complex conjugate to $S$;

2) causality: if $V_1(t)=V_2(t)$ for $t<t_0$, then the operator $S(V_1)*S(V_2)^{-1}$
does not depend on the behavior of the functions $V_1(t)$, $V_2(t)$ at $t<t_0$.

\section{Generalization to higher dimensions}

We are going to generalize the constructions of \S1 to multidimensional space-time $\R^{n+1}$
with coordinates $x=(x^0,\ldots,x^n)$. In this case, instead of configuration space
we have the infinite dimensional Schwartz space of real functions
$\varphi(s)$ on a $n$-dimensional
space-like surface $\CC$ given by smooth parameterization $x=x(s)$, $s=(s^1,\ldots,s^n)$.
Since there are no half-forms and the theory of differential equations is ill-defined [3,4],
it remains to generalize the scattering theory.

We shall restrict ourselves with theory of real scalar field, the generalizations to the
vector case and to fermionic case being straightforward.

\subsection{Hamiltonian formulation of classical field theory} Let us first recall the
covariant Hamiltonian formulation of classical field theory [3,10].

Consider the action functional of the form
\begin{equation}
J=\int L(x,\varphi(x),\varphi_{x^j}(x))dx.
\end{equation}
Then the Euler--Lagrange equations can be written in the following Hamiltonian form.
Introduce the conjugate variables $\pi(s)$ to $\varphi(s)$ by the formula
\begin{equation}
\pi(s)=\sum_l(-1)^lL_{\varphi_{x^l}}
\frac{\partial(x^0,\ldots,\widehat{x^l},\ldots,x^n)}{\partial(s^1,\ldots,s^n)},
\end{equation}
where the fraction means Jacobian.
Introduce also the {\it covariant Hamiltonian densities} $H^j(s)=H^j(x^l(s)$,
$x^l_{s^k}$, $\varphi(s)$, $\varphi_{s^k}$, $\pi(s))$, $j=0,\ldots,n$, by the formulas
\begin{equation}
\begin{aligned}{}
H^j&=\sum_{l\ne j}(-1)^lL_{\varphi_{x^l}}\varphi_{x^j}
\frac{\partial(x^0,\ldots,\widehat{x^l},\ldots,x^n)}{\partial(s^1,\ldots,s^n)}\\
&+(-1)^j(L_{\varphi_{x^j}}\varphi_{x^j}-L)
\frac{\partial(x^0,\ldots,\widehat{x^j},\ldots,x^n)}{\partial(s^1,\ldots,s^n)}.
\end{aligned}
\end{equation}
Then the equations of motion can be written in the form
\begin{equation}
\frac{\delta \Phi(\varphi(\cdot),\pi(\cdot);x^j(\cdot))}{\delta x^j(s)}=\{\Phi,H^j(s)\},
\end{equation}
where $\Phi(\varphi(\cdot),\pi(\cdot);x^j(\cdot))$ is an arbitrary functional of
functions $\varphi(s)$, $\pi(s)$
changing together with the surface $x^j=x^j(s)$, and
\begin{equation}
\{\Phi_1,\Phi_2\}=\int\left(\frac{\delta\Phi_1}{\delta\varphi(s)}
\frac{\delta\Phi_2}{\delta\pi(s)}
-\frac{\delta\Phi_1}{\delta\pi(s)}\frac{\delta\Phi_2}{\delta\varphi(s)}\right)ds
\end{equation}
is the Poisson bracket of two functionals $\Phi_l(\varphi(\cdot),\pi(\cdot))$, $l=1,2$.

\subsection{Definition of the Weyl--Moyal algebra $W_0$}

The Weyl--Moyal algebra of the surface $\CC$ is defined as the algebra of
infinitely differentiable functionals $\Phi(\varphi(\cdot),\pi(\cdot))$ of functions
$\varphi(s)$, $\pi(s)$ from the Schwartz space with respect to
the following (not everywhere defined) {\it Moyal product}:
\begin{equation}
\begin{aligned}{}
&(\Phi_1*\Phi_2)(\varphi(\cdot),\pi(\cdot))\\
&=\exp\left(-\frac{ih}2\int\left(
\frac{\delta}{\delta\varphi_1(s)}
\frac{\delta}{\delta\pi_2(s)}
-\frac{\delta}{\delta\pi_1(s)}\frac{\delta}{\delta\varphi_2(s)}\right)ds\right)\\
&\left.\Phi_1(\varphi_1(\cdot),\pi_1(\cdot))\Phi_2(\varphi_2(\cdot),\pi_2(\cdot))
\right|_{\varphi_1=\varphi_2=\varphi,\pi_1=\pi_2=\pi}.
\end{aligned}
\end{equation}

Consider the case
$$
L(x,\varphi(x),\varphi_{x^j}(x))=L_0(\varphi(x),\varphi_{x^j}(x))
+V(x,\varphi(x),\varphi_{x^j}(x)),
$$
where $L_0$ is quadratic in $\varphi$ and $\varphi_{x^j}$
and independent of $x$,
more concretely,
$$
L_0=\frac12(\varphi_{x^0}^2-\sum_{j=1}^n\varphi_{x^j}^2-m^2\varphi^2),
$$
and $V$ has compact support in $x$. Then {\it the Weyl--Moyal algebra $W_0$} is defined as
the algebra of functionals $\Phi(\varphi(\cdot),\pi(\cdot);x(\cdot))$ with the
Moyal product (10) subject to the following
Heisenberg equation:
\begin{equation}
\frac{\delta\Phi}{\delta x^j(s)}=\frac1{ih}[H_0^j(s),\Phi]=\{\Phi,H_0^j(s)\},
\end{equation}
where $H_0^j(s)$ is the covariant Hamiltonian density corresponding to the Lagrangian $L_0$.
The latter equality in (11) holds because $H_0^j(s)$ is quadratic, and solutions of equation
(11) are well defined just because of this equality.

In fact, the algebra $W_0$ is identified with the Weyl--Moyal algebra of the symplectic vector
space of solutions $\varphi(x)$ of the Klein--Gordon equation
\begin{equation}
\Box\varphi-m^2\varphi\equiv-\frac{\partial^2\varphi}{(\partial x^0)^2}+
\sum_{j=1}^n\frac{\partial^2\varphi}{(\partial x^j)^2}-m^2\varphi=0
\end{equation}
on the whole space-time, i.~e., the algebra of functionals $\Phi(\varphi(\cdot))$
of a solution $\varphi(x)$ with the Moyal product corresponding to the Poisson bracket
on the space of solutions. Below we shall use this realization of the algebra $W_0$.

\subsection{Perturbation theory: the $\varphi^4$ model.} Further on we restrict ourselves by
the typical example of the $\varphi^4$ model in $\R^{3+1}$, i.~e.
$$
V(x,\varphi(x),\varphi_{x^j}(x))=V_0+j(x)\varphi(x),\ \
V_0=\frac{g(x)}{4!}\varphi^4(x),
$$
where $g(x)$ and $j(x)$ are real smooth functions with compact support.
\medskip

{\bf Theorem.}{\it There exists a map from the set of smooth functions $g=g(x), j=j(x)$
with compact support to the set of functionals $S(g,j)\in W_0$ with the following properties.

1) $S(g,j)$ is a formal series in $g,j$ with the first three terms
\begin{equation}
S(g,j)=1+\frac1{ih}\int\left(\frac{g(x)}{4!}\varphi(x)^4+j(x)\varphi(x)\right)dx+\ldots.
\end{equation}

2) Classical limit:
$S(g,j)=a(g,j;h)\exp(iR(g,j)/h)$, where $a(g,j;h)$ is a formal series in $h$, and conjugation
by $\exp(iR(g,j)/h)$ in the Weyl algebra $W_0$ up to $O(h)$ yields the
perturbation series for the evolution operator of the nonlinear classical field equation
\begin{equation}
\Box\varphi(x)-m^2\varphi(x)-g(x)\varphi^3(x)/3!=j(x)
\end{equation}
in the space of functionals $\Phi(\varphi(\cdot))$ from $t=x_0=-\infty$ to $t=\infty$.

3) The Lorentz invariance condition:
\begin{equation}
\Lambda S(\Lambda^{-1}g, \Lambda^{-1}j)=S(g,j)
\end{equation}
for a Lorentz transformation $\Lambda$.

4) The unitarity condition:
\begin{equation}
S(g,j)*\overline{S(g,j)}=1,
\end{equation}
where $\overline{S(g,j)}$ is complex conjugate to $S(g,j)$.

5) The causality condition: for two sets of functions $(g_1,j_1);(g_2,j_2)$ equal for
$t\le t_0$, the product
$S(g_1,j_1)*S(g_2,j_2)^{-1}$ does not depend on the behavior of the functions $g_1,j_1,g_2,j_2$
for $t<t_0$. The same holds for any space-like surface $\CC$ instead of the surface $t=t_0$.

6) The quasiclassical dynamical evolution (cf. with the Maslov--Shvedov quantum field
theory complex germ [5,6]): for any space-like surfaces $\CC_1$, $\CC_2$
there exists a limit $S_{\CC_1,\CC_2}$ of
$S(g,j)$ modulo $o(h)$ as the function $g(x)$ tends to $g=\const$
if $x$ belongs to the strip between
the space-like surfaces and to $0$ otherwise. This limit possesses the property
\begin{equation}
S_{\CC_2,\CC_3}(g)*S_{\CC_1,\CC_2}(g)=S_{\CC_1,\CC_3}(g_1)+o(h),
\end{equation}
where $g_1=g+O(h)$ is a formal series in $g$.

7) The adiabatical interaction switch off:
there exists a limit $S$ of $S(g,j)$ as $g(x)$ tends to the function
$g=\const$. This $S$ is a formal power series in $g$.

Any other choice of $S(g,j)$ with the properties 1--7 above is equivalent to some change of
parameters $g(x)$.
}
\medskip

{\bf Sketch of the proof.} The proof is based on the same ideas as the construction of
Bogolubov $S$-matrix in [2] using the renormalization and the Bogolubov--Parasyuk theorem.
The main difference
with [2] is that instead of algebra of operators in the Fock space one uses the Weyl algebra of
functionals with the Moyal product, instead of the normal ordering of operators in the
Fock space one uses the usual (commutative) product of functionals, and instead of the Feynman
propagator $1/(p^2-m^2+i\varepsilon)$ one uses the function $\PV\,1/(p^2-m^2)$, where
$\PV$ denotes the Cauchy principal value.
\medskip

Similarly to \S1 one introduces the operator Green functions and the
operator Feynman measure $\mu$.

One can conjecture that the scattering operator $S(g,j)\in W_0$ exists outside perturbation
theory as a formal series in $j$, subject to conditions 2--7 of the Theorem.


\begin{thebibliography}{99}
\bibitem{1} A. V. Stoyanovsky, A mathematical interpretation of the Feynman path integral
equivalent to the formalism of Green functions, arxiv.org: 0808.1511 [math-ph], submitted to
Math. Zametki.
\bibitem{2} N. N. Bogolubov and D. V. Shirkov, Introduction to the theory of quantized
fields, Moscow, GITTL, 1957 (in Russian).
\bibitem{3} A. V. Stoyanovsky, Introduction to the mathematical principles of quantum field
theory, Moscow, Editorial URSS, 2007 (in Russian).
\bibitem{4} A. V. Stoyanovsky,
Maslov's complex germ and the Weyl--Moyal algebra in quantum mechanics
and in quantum field theory, math-ph/0702094, to appear in Russian Journal of Math. Phys.
\bibitem{5} V. P. Maslov, O. Yu. Shvedov, Method of complex germ in the many-body problem
and in quantum field theory, Moscow, Editorial URSS, 2000 (in Russian).
\bibitem{6} A. V. Stoyanovsky, A necessary condition for existence of S-matrix outside
perturbation theory, arxiv:0707.4570 [hep-th], Mat. Zametki, vol. 83, No. 4, 613--617, 2008.
\bibitem{7} V. P. Maslov, Quantization of thermodynamics and ultrasecondary quantization,
Moscow, Institute for Computer Studies, 2001 (in Russian).
\bibitem{8} A. V. Stoyanovsky, Pseudodynamical evolution and path integral in quantum field
theory, arxiv.org:0810.4854 [math-ph], to appear in Doklady Mathematics.
\bibitem{9} L. Hormander, The analysis of linear partial differential operators, vol. III,
Springer Verlag, 1985.
\bibitem{10} A. V. Stoyanovsky, Generalized Schrodinger equation for free field, hep-th/0601080.
\end{thebibliography}
\end{document}